\documentclass[prb, superscriptaddress, twocolumn]{revtex4}

\usepackage{amsmath, amssymb, braket}
\usepackage{mathrsfs}
\usepackage{epstopdf}
\usepackage{graphicx, pdflscape, array, dcolumn, makecell}
\usepackage[sort&compress]{natbib}
\usepackage{natmove, mciteplus}
\usepackage[pdftex = true, colorlinks = true, citecolor = blue, bookmarks = true]{hyperref}
\usepackage{hypernat}
\usepackage{bibentry}
\usepackage[utf8]{inputenc}
\usepackage[OT4]{fontenc}

\newcommand{\beq}{\begin{equation}}

\newcommand{\eeq}{\end{equation}}

\newcommand{\bec}{\begin{center}}

\newcommand{\eec}{\end{center}}

\begin{document}


\title{Density Functional Theory Approach to Noncovalent Interactions \\ via Interacting Monomer Densities}
\date{\today}
\author{Łukasz Rajchel}
\email{lrajchel@tiger.chem.uw.edu.pl}
\affiliation{Department of Chemistry, Oakland University, Rochester, Michigan 48309-4477, USA}
\affiliation{Faculty of Chemistry, University of Warsaw, 02-093 Warszawa, Pasteura 1, Poland}
\author{Piotr S. Żuchowski}
\affiliation{Department of Chemistry, Durham University, South Road, Durham DH1 3LE, United Kingdom}
\author{Michał Hapka}
\affiliation{Faculty of Chemistry, University of Warsaw, 02-093 Warszawa, Pasteura 1, Poland}
\author{Marcin~Modrzejewski}
\affiliation{Faculty of Chemistry, University of Warsaw, 02-093 Warszawa, Pasteura 1, Poland}
\author{Małgorzata M. Szczęśniak}
\affiliation{Department of Chemistry, Oakland University, Rochester, Michigan 48309-4477, USA}
\author{Grzegorz Chałasiński}
\email{chalbie@tiger.chem.uw.edu.pl}
\affiliation{Faculty of Chemistry, University of Warsaw, 02-093 Warszawa, Pasteura 1, Poland}


\begin{abstract}
A recently proposed "DFT+dispersion" treatment (\citeauthor{rajchel_zuch_szczesniak_chal:2010_prl}, \textit{Phys. Rev. Lett.}, \citeyear{rajchel_zuch_szczesniak_chal:2010_prl}, \textbf{104}, 163001) is described in detail and illustrated by more examples. The formalism derives the dispersion-free density functional theory (DFT) interaction energy and combines it with the dispersion energy from separate DFT calculations. It consists in the self-consistent polarization of DFT monomers restrained by the exclusion principle via the Pauli blockade technique. Within the monomers a complete exchange-correlation potential should be used, but between them only the exact exchange operates. The applications to wide range of molecular complexes from rare-gas dimers to H-bonds to $\pi$-electron interactions show good agreement with benchmark values.
\end{abstract}

\maketitle


\section{Introduction}

Density Functional Theory~(DFT)-based methods provide the most important viable approach to large systems of nano- and biotechnological relevance. Their success for determination of structure, energetics and other static properties of dense matter is well known~\cite{koch_holt:2001, parr_yang:1989, perdew_kurt:2003}.

However, treatment of weak non-covalent interactions by DFT remains plagued by spurious and erratic results. This is a consequence of the fact that stabilization in these complexes is determined by dispersion interaction, not accounted for in standard DFT functionals~\cite{tsuzuki_luthi:2001, wu_vargas_nayak_lot_scoles:2001}. Consequently, exchange-correlation potentials derived from local and semi-local models and combined to satisfy empirical data often feature artifacts when applied to systems with large non-local correlation effects.

Serious efforts have recently been invested to incorporate the dispersion effect into the DFT framework, and the results are promising~\cite{lund_ander_shao_chan_lang:1995, grimme:2004, chai_head-gordon:2008, lilien_tavern_roth_sebas:2004, truhlar_zhao:2006, toul_gerber_jansen_savin_angyan:2009, janesko_hend_scus:2009}. At the same time, in many practical applications remarkable progress has been achieved by using a posteriori dispersion corrections, model and/or semi-empirical, added on the top of regular DFT calculations~(DFT+D)~\cite{grimme:2004} in the spirit of the classic SCF+dispersion model of~\citet{ahlr_penco_scoles:1977} and~\citet{wu_vargas_nayak_lot_scoles:2001} 

In order to make a DFT+dispersion strategy successful, one needs two ingredients: a dispersion-free DFT interaction energy~\cite{rajchel_zuch_szczesniak_chal:2010_prl} and the model for the dispersion energy. The first, the DFT dispersion component, may be obtained via the SAPT approach, but also from other approximate DFT treatments~\cite{lund_ander_shao_chan_lang:1995, grimme:2004, neumann_perrin:2005, becke_johnson:2007}. The second ingredient, the dispersion-free DFT interaction energy, is commonly obtained  as the supermolecular DFT interaction energy. However, in contrast to the SCF interaction energy, the composition of the DFT interaction energy is neither understood nor controlled. In particular, an exchange-correlation functional is always an approximation  to some extent delocalized, and never exactly dispersion-free, and the problem of double counting of obscure dispersion terms arises. In addition, the exchange-correlation potential compromises the demands of many users and various training sets, and thus is prone to erroneous behavior~\cite{john_becke_sherr_diliabio:2009}. 

A rigorous approach requires a DFT interaction energy that a priori neglects non-local long-range interaction energy terms (dispersion) but allows for accurate mutual exchange and mutual polarization effects --- an analogue of the SCF interaction energy at the DFT level of theory. Such a DFT interaction energy could be confidently and rigorously supplemented with a dispersion component obtained also at the DFT level of theory using SAPT~\cite{mis_jez_szal:2003, hess_jansen:2003} or other formalisms~\cite{lund_ander_shao_chan_lang:1995, grimme:2004, becke_johnson:2007}. The goal of this work has been to define an accurate DFT+dispersion treatment which is based the derivation of "dispersion-free" interaction energy arising between DFT monomers to which a posteriori DFT dispersion energy is added. The former, the dispersion-free DFT interaction energy, leaves out any intermolecular correlation terms --- thus avoiding artifactual exchange and doubly-counted dispersion terms. The latter may be obtained from e.g. TDDFT and/or SAPT calculations. The basic formulation of the method and selected results for model systems: rare-gas dimers and H-bonded systems, have already been presented by~\citet{rajchel_zuch_szczesniak_chal:2010_prl}.

To this end a novel hybrid DFT approach for calculations of van der Waals complexes has been proposed by~\citet{rajchel_zuch_szczesniak_chal:2010_prl}. It uses the formalism of~\citet{gut_pauli:1988} termed Pauli blockade-Hartree Fock~(PB-HF) combined with the bifunctional formulation of DFT of~\citet{rajchel_zuch_szczesniak_chal:2010_cpl}. At the first stage, the subsystems' sets of orbitals are separated by symmetrical orthogonalization of the orbitals residing at different monomers. At the second stage, one iteratively evaluates the interaction energy between two DFT monomers described by Kohn-Sham determinants in a manner analogous to the HF method. That is, the monomers are mutually polarized until self-consistency in each other fields is reached under the constraint of the Pauli exclusion principle (Pauli blockade technique). Various exchange-correlation DFT potentials may be used within the monomers (see below), but the intermonomer exchange-correlation potential is reduced to only the exact exchange, and thus neglects the dispersion contribution. In the third stage, the dispersion component is a posteriori added, from SAPT(DFT) calculations. In such a way, the erratic behavior of approximate exchange functionals around the equilibrium separations is no longer the case, while the missing long-range attraction caused by dispersion is added without the problem of double counting.

It is worthwhile to note that the above approach bears resemblance to the range-separation idea in DFT~\cite{tawada_tsuneda_yanag_yanai_hirao:2004, angyan_gerber_savin_toul:2005, peach_helgaker_salek_keal_lutnaes_tozer_handy:2006} differing in that it is based on separation of monomers rather than ranges.

In Sec.~\ref{sec:theory} the "dispersion-free" PB(DFT) is derived as a particular case of the PB treatment of DFT reported by us recently.~\cite{rajchel_zuch_szczesniak_chal:2010_cpl} In Sec~\ref{sec:res_disc} we report numerical results of our method for representative van der Waals and H-bonded systems, and discuss the overall performance of the method. 

\section{Theory}
\label{sec:theory}

\subsection{Pauli blockade method}

Supermolecular energy in terms of DFT can be defined as the difference between the total energies of the dimer AB and the individual monomers A and B, separated to infinity:
	\beq
		E_\mathrm{int}^\mathrm{DFT} = E_\mathrm{AB}^\mathrm{DFT} - E_\mathrm{A}^\mathrm{DFT} - E_\mathrm{B}^\mathrm{DFT}.
		\label{eq:PB_DFTint}
	\eeq
The interaction energy calculated in this way contains the correlation contribution, however the long-range correlation effects are not taken into account correctly within standard density functionals.

It was demonstrated by \citet{gut_pauli:1988} that the Hartree-Fock~(HF) supermolecular interaction energy may be exactly recovered by solving the HF equations for \emph{monomers} in the presence of the external perturbation consisting of the electrostatic potential and the non-local exchange potential generated by the second monomer. They have also proposed a convenient computational scheme in terms of mutually orthogonalized A~and B~monomers' occupied orbitals, $\left\{ \tilde{a}_i \right\}_{i \in A}$ and $\left\{ \tilde{b}_k \right\}_{k \in B}$ (the quantities expressed in the orthonormalized orbitals are henceforth marked with tilde). With such orbitals, satisfying
	\beq
		\forall_{i \in A} \forall_{k \in B}: \Braket{ \tilde{a}_i | \tilde{b}_k } = 0,
		\label{eq:PB_tab_orth}
	\eeq
the supermolecular interaction energy in Hartree-Fock method can be written as
	\beq
		E_\mathrm{int}^\mathrm{HF} = \Delta \tilde{E}_\mathrm{A} + \Delta \tilde{E}_\mathrm{B} + \tilde{E}_\mathrm{elst} + \tilde{E}_\mathrm{exch},
	\eeq
where
	\beq
		\Delta \tilde{E}_\mathrm{A} = \tilde{E}_\mathrm{A}^\mathrm{HF} - E_\mathrm{A}^\mathrm{HF}
	\eeq
is the difference between the final and isolated monomer~A HF energy,
	\begin{align}
		& \tilde{E}_\mathrm{elst} =
			\int_{\mathbb{R}^3} v_\mathrm{A}^\mathrm{ne}(\mathbf{r}) \tilde{\rho}_\mathrm{B}(\mathbf{r}) \, d^3 \mathbf{r} +
			\int_{\mathbb{R}^3} v_\mathrm{B}^\mathrm{ne}(\mathbf{r}) \tilde{\rho}_\mathrm{A}(\mathbf{r}) \, d^3 \mathbf{r} + \nonumber \\
		& + 4 \sum_{i \in A} \sum_{k \in B} \Braket{ \tilde{a}_i \tilde{b}_k | \tilde{a}_i \tilde{b}_k } + V_\mathrm{int}^\mathrm{nn},
		\label{eq:PB_Eelst}
	\end{align}
is the electrostatic interaction,
	\beq
		\tilde{\rho}_\mathrm{A}(\mathbf{r}) = 2 \sum_{i \in A} \left| \tilde{a}_i(\mathbf{r}) \right|^2
		\label{eq:PB_densA}
	\eeq
is the monomer~A density,
	\beq
		v_\mathrm{A}^\mathrm{ne}(\mathbf{r}) = - \sum_{\alpha \in \mathcal{A}} \frac{Z_\alpha}{|\mathbf{r} - \mathbf{R}_\alpha|}
	\eeq
is the monomer~A nuclear potential with~$\alpha$ labeling the coordinates of monomer~A's nuclei, each described by its position~$\mathbf{R}_\alpha$ and charge~$Z_\alpha$,
	\beq
		V_\mathrm{int}^\mathrm{nn} = \sum_{\alpha \in \mathcal{A}} \sum_{\beta \in \mathcal{B}} \frac{Z_\alpha Z_\beta}{|\mathbf{R}_\alpha - \mathbf{R}_\beta|}
	\eeq	
is the intermonomer nuclear repulsion energy (constant for a fixed geometry), and finally
	\beq
		\tilde{E}_\mathrm{exch} = -2 \sum_{i \in A} \sum_{k \in B} \Braket{ \tilde{a}_i \tilde{b}_k | \tilde{b}_k \tilde{a}_i },
		\label{eq:PB_Eexch}
	\eeq
is the exchange interaction. The monomer~A orbitals are the eigenfunctions of the following modified Fock operator:
	\beq
		\hat{\tilde{f}}_\mathrm{A} + \hat{\tilde{v}}_\mathrm{B}^\mathrm{elst} + \hat{\tilde{v}}_\mathrm{B}^\mathrm{exch},
		\label{eq:PB_modFock}
	\eeq
where $\hat{\tilde{f}}_\mathrm{A}$ is the standard Fock operator build of $\left\{ \tilde{a}_i \right\}_{i \in A}$ orbital set, and the two remaining terms are electrostatic and the non-local exchange potentials generated by monomer~B, respectively. Monomer~B orbitals are obtained analogously. 

The procedure introduced by \citet{gut_pauli:1988} can be generalized to the case where the two subsystems are described by KS orbitals. To this end we note that the total density of the system can be represented as the sum of monomer densities obtained from orthogonalized orbitals:
	\beq
		\rho_\mathrm{AB} = \tilde{\rho}_\mathrm{AB} = \tilde{\rho}_\mathrm{A} + \tilde{\rho}_\mathrm{B},
		\label{eq:PB_trhoAB_add}
	\eeq
where the monomer~A density is calculated as in~\eqref{eq:PB_densA} but with orbitals being the solutions of the following modified KS equation:
	\beq
		\Big( \hat{\tilde{f}}_\mathrm{A}^\mathrm{KS}(\mathbf{r}) + \Delta \tilde{v}_\mathrm{A}^\mathrm{xc}(\mathbf{r}) + \hat{\tilde{v}}_\mathrm{B}^\mathrm{elst}(\mathbf{r}) \Big) \tilde{a}_i(\mathbf{r}) = \epsilon_{\mathrm{A}, i} \tilde{a}_i(\mathbf{r})
		\label{eq:PB_KSeqA}
	\eeq
and satisfying~\eqref{eq:PB_tab_orth}. In~Eq.~\eqref{eq:PB_KSeqA}, $\hat{\tilde{f}}_\mathrm{A}^\mathrm{KS}$ is the standard KS operator built of $\left\{ \tilde{a}_i \right\}_{i \in A}$ orbitals,
	\beq
		\Delta \tilde{v}_\mathrm{A}^\mathrm{xc}(\mathbf{r}) = v_\mathrm{AB}^\mathrm{xc}(\mathbf{r}) - \tilde{v}_\mathrm{A}^\mathrm{xc}(\mathbf{r}),
		\label{eq:PB_Dvxc}
	\eeq
is the non-additivity of the monomer~A exchange-correlation~(xc) potential, and $\hat{\tilde{v}}_\mathrm{B}^\mathrm{elst}(\mathbf{r})$ is the electrostatic potential of~Eq.~\eqref{eq:PB_modFock}. With $\rho_\mathrm{AB}$ decomposed in such way, the total KS interaction energy may be written as a functional of~$\tilde{\rho}_\mathrm{A}$ and~$\tilde{\rho}_\mathrm{B}$ densities:
	\begin{align}
		& E_\mathrm{int}^\mathrm{PB} \left[ \tilde{\rho}_\mathrm{A}; \tilde{\rho}_\mathrm{B} \right] =
			\Delta \tilde{E}_\mathrm{A} \left[ \tilde{\rho}_\mathrm{A} \right] + \Delta \tilde{E}_\mathrm{B} \left[ \tilde{\rho}_\mathrm{B} \right] + \nonumber \\
		& + E_\mathrm{elst} \left[ \tilde{\rho}_\mathrm{A}; \tilde{\rho}_\mathrm{B} \right] +
			\Delta E_{\mathrm{xc}} \left[ \tilde{\rho}_\mathrm{A}; \tilde{\rho}_\mathrm{B} \right].
		\label{eq:PB_Eint_bifun}
	\end{align}
The terms of~\eqref{eq:PB_Eint_bifun} are as follows. Monomer~A deformation is
	\beq
		\Delta \tilde{E}_\mathrm{A} \left[ \tilde{\rho}_\mathrm{A} \right] = E_\mathrm{A} \left[ \tilde{\rho}_\mathrm{A} \right] - E_\mathrm{A} \left[ \rho^0_\mathrm{A} \right],
	\eeq
and~$\rho^0_\mathrm{A}$ is the density of the unperturbed A~monomer, i.e. the density built of orbitals~$\left\{ a_i^0 \right\}_{i \in A}$ satisfying unperturbed KS equations,
	\beq
		\hat{f}^{\mathrm{KS},0}(\mathbf{r}) a_i^0(\mathbf{r}) = \epsilon_{\mathrm{A}, i}^0 a_i^0(\mathbf{r}).
		\label{eq:PB_a0}
	\eeq
The total energy of monomer~A is expressed via the standard KS functional,
	\begin{align}
		& E_\mathrm{A} \left[ \tilde{\rho}_\mathrm{A} \right] = T^\mathrm{s} \left[ \tilde{\rho}_\mathrm{A} \right] + V_\mathrm{A}^\mathrm{ne} \left[ \tilde{\rho}_\mathrm{A} \right] +
			J \left[ \tilde{\rho}_\mathrm{A} \right] + \nonumber \\
		& + E^\mathrm{xc} \left[ \tilde{\rho}_\mathrm{A} \right] + V_\mathrm{A}^\mathrm{nn}.
		\label{eq:PB_fundef_TsVneJExc}
	\end{align}
The functional~\eqref{eq:PB_fundef_TsVneJExc} includes the non-iteracting kinetic energy:
	\beq
		T^\mathrm{s} \left[ \tilde{\rho}_\mathrm{A} \right] = -\sum_{i \in A} \Braket{ \tilde{a}_i | \Delta_\mathbf{r} | \tilde{a}_i },
		\label{eq:PB_Ts_fun}
	\eeq
nuclear-electron attraction energy:
	\beq
		V_\mathrm{A}^\mathrm{ne} \left[ \tilde{\rho}_\mathrm{A} \right] = \int_{\mathbb{R}^3} v_\mathrm{A}^\mathrm{ne}(\mathbf{r}) \tilde{\rho}_\mathrm{A}(\mathbf{r}) \, d^3 \mathbf{r},
	\eeq
coulombic energy:
	\beq
		J \left[ \tilde{\rho}_\mathrm{A} \right] =
			\frac{1}{2} \int_{\mathbb{R}^3} \int_{\mathbb{R}^3} \frac{\tilde{\rho}_\mathrm{A}(\mathbf{r}_1) \tilde{\rho}_\mathrm{A}(\mathbf{r}_2)}{r_{12}} \, d^3 \mathbf{r}_1 d^3 \mathbf{r}_2,
	\eeq
exchange-correlation~(xc) energy:
	\beq
		E^\mathrm{xc} \left[ \tilde{\rho}_\mathrm{A} \right] =
			\int_{\mathbb{R}^3} F^\mathrm{xc} \Big( \tilde{\rho}_\mathrm{A}(\mathbf{r}); \Big\{ \nabla_\mathbf{r} \tilde{\rho}_\mathrm{A}(\mathbf{r}); \ldots \Big\} \Big) \, d^3\mathbf{r}
	\eeq
which is evaluated through the numerical integration of the~$F^\mathrm{xc}$ integrand on a~grid of points around monomer~A. The last term of~\eqref{eq:PB_fundef_TsVneJExc} is the monomer~A nuclear-nuclear repulsion energy. Similar expressions can be written for monomer~B. The electrostatic part of the interaction energy has the same form as in Eq.~\eqref{eq:PB_Eelst} and in terms of densities it can be easily rewritten as
	\begin{align}
		& E_\mathrm{elst}[\tilde{\rho}_\mathrm{A}; \tilde{\rho}_\mathrm{B}] = \nonumber \\
		& = \int_{\mathbb{R}^3} v_\mathrm{B}^\mathrm{ne}(\mathbf{r}) \tilde{\rho}_\mathrm{A}(\mathbf{r}) \, d^3 \mathbf{r} +
			\int_{\mathbb{R}^3} v_\mathrm{A}^\mathrm{ne}(\mathbf{r}) \tilde{\rho}_\mathrm{B}(\mathbf{r}) \, d^3 \mathbf{r} + \nonumber \\
		& + \int_{\mathbb{R}^3} \int_{\mathbb{R}^3} \frac{\tilde{\rho}_\mathrm{A}(\mathbf{r}_1) \tilde{\rho}_\mathrm{B}(\mathbf{r}_2)}{r_{12}} \, d^3 \mathbf{r}_1 d^3 \mathbf{r}_2 +
			V_\mathrm{int}^\mathrm{nn}.
		\label{eq:PB_tEelst}
	\end{align}
Finally, the exchange-correlation interaction is calculated in a~supermolecular manner,
	\beq
		\Delta E_\mathrm{xc}[\tilde{\rho}_\mathrm{A}; \tilde{\rho}_\mathrm{B}] = E^\mathrm{xc}[\tilde{\rho}_\mathrm{A} + \tilde{\rho}_\mathrm{B}] -
			E^\mathrm{xc}[\tilde{\rho}_\mathrm{A}] - E^\mathrm{xc}[\tilde{\rho}_\mathrm{B}].
		\label{eq:PB_Excint}
	\eeq
Inserting monomer densities calculated using~Eq.~\eqref{eq:PB_densA} with orbitals satisfying~\eqref{eq:PB_tab_orth} and~\eqref{eq:PB_KSeqA} into~\eqref{eq:PB_Eint_bifun} one restores the DFT supermolecular interaction energy~\eqref{eq:PB_DFTint}. The expression~\eqref{eq:PB_Eint_bifun} is potentially exact, i.e. it yields the exact interaction energy provided that the exact xc potential is used.

Technically, the orbitals satisfying~\eqref{eq:PB_KSeqA} are found in a self-consistent iterative process and the orthogonality condition~\eqref{eq:PB_tab_orth} is imposed by the brute-force incorporation of the penalty operator and successive Löwdin orthogonalization. Depicting iteration numbers in square brackets, the $n$th iterative step for monomer~A reads
	\begin{align}
		& \Big( \hat{\tilde{f}}_\mathrm{A}^{\mathrm{KS}[n - 1]} + \Delta \hat{\tilde{v}}_\mathrm{A}^{\mathrm{xc}[n - 1]} + \hat{\tilde{v}}_\mathrm{B}^{\mathrm{elst}[n - 1]} +
			\eta \hat{\tilde{R}}_\mathrm{B}^{[n - 1]} \Big) a_i^{[n]} = \nonumber \\
		& = \epsilon_{\mathrm{A}, i}^{[n]} a_i^{[n]},
		\label{eq:PB_full_SC_ab_iter}
	\end{align}
where the penalty operator is
	\beq
		\hat{\tilde{R}}_\mathrm{B}^{[n]} = \sum_{k \in B} \Ket{\tilde{b}_k^{[n]}} \Bra{\tilde{b}_k^{[n]}}.
		\label{eq:PB_penfun}
	\eeq
and it is obvious that its action on monomer~A's occupied orbitals annihilates them once the orbitals are orthogonal. $\eta > 0$ is a~scaling parameter not affecting the final solutions. The equivalent of~\eqref{eq:PB_full_SC_ab_iter} for monomer~B is obtained through the interchange of the A and~B subscripts. The orbitals obtained in each iteration are orthogonalized, yielding an orthonormal
	\beq
		\left\{ \left\{ \tilde{a}_i^{[n]} \right\}_{i \in A}; \left\{ \tilde{b}_k^{[n]} \right\}_{k \in B} \right\}
		\label{eq:PB_orth_set}
	\eeq
set. The iterations start with the unperturbed orbitals obtained in~Eq.~\eqref{eq:PB_a0} and its analogue for monomer~B.

The zeroth order interaction energy may be viewed as an analog of the well-known HF-based Heitler-London~(HL) interaction energy. Specifically, we define the DFT-based HL interaction energy as
	\beq
		E_\mathrm{int}^\mathrm{HL} = E_\mathrm{AB}\left[ \tilde{\rho}_\mathrm{A}^0; \tilde{\rho}_\mathrm{B}^0 \right] - E \left[ \rho^0_\mathrm{A} \right] - E \left[ \rho^0_\mathrm{B} \right],
		\label{eq:PB_Eint_HL}
	\eeq
where the densities $\tilde{\rho}_\mathrm{A}^0$~and~$\tilde{\rho}_\mathrm{B}^0$ are obtained as in~\eqref{eq:PB_densA} from the orbitals generated through the orthogonalization of the unperturbed orbitals~$\left\{ \left\{ a_i^0 \right\}_{i \in A}; \left\{ b_k^0 \right\}_{k \in B} \right\}$.
This definition is equivalent to that proposed by~\citet{cyb_seversen:2003}.

For the proof and the detailed discussion of DFT based PB method the Reader is referred elsewhere~\cite{rajchel_zuch_szczesniak_chal:2010_cpl}.

\subsection{Dispersion-free approximation}

The Eq.~\eqref{eq:PB_Eint_bifun} shows how the DFT interaction energy can be evaluated without refering to supermolecule concept, using exclusively appropriately perturbed KS equations solved for monomers. The decomposition of the DFT interaction energy allows to modify it in such a way that we can eliminate the correlation effects between two subsystems, so that we obtain the dispersionless interaction energy between two systems. To this end, we describe the interacting monomers with the full xc~potentials while allowing only the exact exchange potential operate between them. Mathematically, this involves the replacement of the $\Delta E_\mathrm{xc}$ term~\eqref{eq:PB_Excint} in~\eqref{eq:PB_Eint_bifun} with the exchange interaction which in terms of one-electron reduced density matrices~(1-DM) for the closed-shell system reads (cf.~Ref.~\onlinecite{mcweeny:1989})
	\begin{align}
		& E_\mathrm{exch}[\tilde{\rho}_\mathrm{A}; \tilde{\rho}_\mathrm{B}] = \nonumber \\
		& = -\frac{1}{2} \int_{\mathbb{R}^3} \int_{\mathbb{R}^3} \frac{\tilde{\rho}_\mathrm{A}(\mathbf{r}_1; \mathbf{r}_2) \tilde{\rho}_\mathrm{B}(\mathbf{r}_2; \mathbf{r}_1)}{r_{12}} \, d^3 \mathbf{r}_1 d^3 \mathbf{r}_2.
	\end{align}
The 1-DM resulting from a~single Slater determinant is
	\beq
		\tilde{\rho}_\mathrm{A}(\mathbf{r}_1; \mathbf{r}_2) = 2 \sum_{i \in A} \tilde{a}_i(\mathbf{r}_1) \tilde{a}^*_i(\mathbf{r}_2)
	\eeq
with the~density~\eqref{eq:PB_densA} simply being the diagonal part of 1-DM,
	\beq
		\tilde{\rho}_\mathrm{A}(\mathbf{r}) \equiv \tilde{\rho}_\mathrm{A}(\mathbf{r}; \mathbf{r}).
	\eeq
Thus, the dimer energy bifunctional incorporating the exact exchange takes the form
	\begin{align}
		& \mathscr{E}_\mathrm{AB}[\tilde{\rho}_\mathrm{A}; \tilde{\rho}_\mathrm{B}] =
			E_\mathrm{A}[\tilde{\rho}_\mathrm{A}] + E_\mathrm{B}[\tilde{\rho}_\mathrm{B}] + \nonumber \\
		& + E_\mathrm{elst}[\tilde{\rho}_\mathrm{A}; \tilde{\rho}_\mathrm{B}] + E_\mathrm{exch}[\tilde{\rho}_\mathrm{A}; \tilde{\rho}_\mathrm{B}].
		\label{eq:PB_EAB_bifun_exexch}
	\end{align}
The idea of reducing the intermolecular potential to the exchange-only part has been successfully used in the model helium dimer calculations by~\citet{hess_jans_pccp:2003} and \citet{allen_tozer:2002}. Now we perform the search of the extremals of the bifunctional~\eqref{eq:PB_EAB_bifun_exexch} with respect to~$\tilde{\rho}_\mathrm{A}$~and~$\tilde{\rho}_\mathrm{B}$ under the constraint of the mutual orthogonality between monomers' occupied orbitals. The orthogonality constraint ensures that the density additivity condition~\eqref{eq:PB_trhoAB_add} is maintained and the intersystem Pauli exclusion principle is fulfilled. To this end, we perform the variational optimization in two steps using the~PB method~(see Refs.~\onlinecite{gut_pauli:1988} and~\onlinecite{rajchel_zuch_szczesniak_chal:2010_cpl}): first, the bifunctional extremal search is performed without the imposition of the intermonomer orthogonality constraint, and secondly, the penalty operator is added in the resulting iterative scheme. The minimization of the bifunctional~\eqref{eq:PB_EAB_bifun_exexch} leads to a system of coupled equations for optimum orbitals:
		\beq
		\begin{cases}
			\Big( \hat{\tilde{f}}_\mathrm{A}^\mathrm{KS}(\mathbf{r}) + \hat{\tilde{v}}_\mathrm{B}^\mathrm{elst}(\mathbf{r}) + \hat{\tilde{v}}_\mathrm{B}^\mathrm{exch}(\mathbf{r}) \Big) \tilde{a}_i(\mathbf{r}) = \epsilon_{\mathrm{A}, i} \tilde{a}_i(\mathbf{r}) \\
			\Big( \hat{\tilde{f}}_\mathrm{B}^\mathrm{KS}(\mathbf{r}) + \hat{\tilde{v}}_\mathrm{A}^\mathrm{elst}(\mathbf{r}) + \hat{\tilde{v}}_\mathrm{A}^\mathrm{exch}(\mathbf{r}) \Big) \tilde{b}_k(\mathbf{r}) = \epsilon_{\mathrm{B}, k} \tilde{b}_k(\mathbf{r})
		\end{cases},
		\label{eq:PB_dfree_SC_ab}
	\eeq
where the action of the exchange operator on an arbitrary one-electron function~$x$ reads
	\beq
		\hat{\tilde{v}}_\mathrm{A}^\mathrm{exch}(\mathbf{r}) x(\mathbf{r}) =
			-\frac{1}{2} \int_{\mathbb{R}^3} \frac{\tilde{\rho}_\mathrm{A}(\mathbf{r}; \mathbf{r}')}{|\mathbf{r} - \mathbf{r}'|} x(\mathbf{r}') \, d^3 \mathbf{r}'.
	\eeq
In the second step of the PB procedure, the iterative process of solving Eqs.~\eqref{eq:PB_dfree_SC_ab} with the aid of the penalty operator is formulated in full analogy to~\eqref{eq:PB_full_SC_ab_iter}. The interaction energy at the $n$th iteration is obtained upon the insertion of the densities calculated with orthogonalized orbitals resulting from~\eqref{eq:PB_orth_set} into~\eqref{eq:PB_EAB_bifun_exexch} and subtracting the unperturbed monomer energies:
	\begin{align}
		& \mathscr{E}_\mathrm{int}^{\mathrm{PB}[n]} = \mathscr{E}_\mathrm{AB} \left[ \tilde{\rho}_\mathrm{A}^{[n]}; \tilde{\rho}_\mathrm{B}^{[n]} \right]
			- E_\mathrm{A} \left[ \rho^0_\mathrm{A} \right] - E_\mathrm{B} \left[ \rho^0_\mathrm{B} \right] = \nonumber \\
		& = \Delta \tilde{E}_\mathrm{A}^{[n]} + \Delta \tilde{E}_\mathrm{B}^{[n]} +
			E_\mathrm{elst} \left[ \tilde{\rho}_\mathrm{A}^{[n]}; \tilde{\rho}_\mathrm{B}^{[n]} \right] +
			E_\mathrm{exch} \left[ \tilde{\rho}_\mathrm{A}^{[n]}; \tilde{\rho}_\mathrm{B}^{[n]} \right].
		\label{eq:PB_Eint_bifun_dfree}
	\end{align}
In the above equation, the A~monomer deformation is
	\beq
		\Delta \tilde{E}_\mathrm{A} = E_\mathrm{A} \left[ \tilde{\rho}_\mathrm{A}^{[n]} \right] - E_\mathrm{A} \left[ \rho^0_\mathrm{A} \right],
	\eeq
and analogously for monomer~B. Interaction energy at the zero iteration is then
	\beq
		\mathscr{E}_\mathrm{int}^\mathrm{HL} = \mathscr{E}_\mathrm{AB} \left[ \tilde{\rho}_\mathrm{A}^0; \tilde{\rho}_\mathrm{B}^0 \right] -
			E \left[ \rho^0_\mathrm{A} \right] - E \left[ \rho^0_\mathrm{B} \right].
		\label{eq:PB_Eint_HL_exexch}
	\eeq
Henceforth, the energy~\eqref{eq:PB_Eint_HL_exexch} will be referred to as the dispersion-free HL interaction energy. The definition of the analog of HL energy at the DFT level~\eqref{eq:PB_Eint_HL}, which is simply the zero iteration obtained with monomer densities unperturbed by the interaction, depends on a particular functional, as shown by~\citet{cyb_seversen:2003} and by~\citet{rajchel_zuch_szczesniak_chal:2010_cpl} The dispersion-free HL interaction energy~\eqref{eq:PB_Eint_HL_exexch}, in contrast to other definitions, rigorously excludes the dispersion interaction. It contains the well defined electrostatic and exchange interaction contributions, in this case between unperturbed isolated DFT monomers. It is closely related to the first order energy in the symmetry-based perturbation theory based on SAPT(DFT).

$\mathscr{E}_\mathrm{int}^\mathrm{PB}$ represents the final PB energy calculated with~\eqref{eq:PB_Eint_bifun_dfree} using self-consistent orbitals satisfying~\eqref{eq:PB_dfree_SC_ab}. It results from the mutual electric polarization of DFT monomers, and, owing to the Pauli blockade procedure, it contains exchange contributions. It is related to the induction terms of the SAPT formalism with two important advantages over the latter: PB sums all electric polarization terms to infinity and accounts for accompanying exchange effects in a consistent manner within the DFT formalism.

In the original HF-based formulation~\cite{gut_pauli:1988}, the~PB procedure simply restores supermolecular HF interaction energy, and obviously neglects any kind of electron correlation. For the DFT analog, both monomers are described with the full KS operator, but are coupled using HF Coulomb and exchange operators~[$\hat{\tilde{v}}_\mathrm{B}^{\mathrm{elst}}$ and~$\hat{\tilde{v}}_\mathrm{B}^{\mathrm{exch}}$ in~Eq.~\eqref{eq:PB_dfree_SC_ab}, respectively] built from KS orbitals. Such an approach accounts for intramonomer local electron correlation leaving out the intermonomer nonlocal contributions. The~$\mathscr{E}_\mathrm{int}^\mathrm{PB}$ represents then "non-dispersion" part of the interaction energy that includes the electrostatic, exchange, and induction components. For rare gas dimers it is purely repulsive~(see section~\ref{sec:res_disc}).

\subsection{Pauli Blockade Plus Dispersion}

The total interaction energy, termed PBD for Pauli blockade plus dispersion, is obtained by adding to~$\mathscr{E}_\mathrm{int}^\mathrm{PB}$ the dispersion component obtained at the DFT level of theory, either from SAPT or by other accurate techniques:
	\beq	
		E_\mathrm{int}^\mathrm{PBD} = \mathscr{E}_\mathrm{int}^\mathrm{PB} + E_\mathrm{disp}.
		\label{eq:PBD_int}
	\eeq
In this work we used the second-order dispersion components from SAPT-DFT~\cite{mis_pod_jez_szal:2005, jans_hess:2001}.
	\beq	
		E_\mathrm{int}^\mathrm{PBD} = \mathscr{E}_\mathrm{int}^\mathrm{PB} + E_\mathrm{disp}^{(2)} + E_\mathrm{exch\text{-}disp}^{(2)}.
		\label{eq:PBD_int}
	\eeq
For the sake of comparison we also calculated the SAPT interaction energies by its most efficient hybrid version, SAPT$\delta$, i.e. through the second order and corrected with the so called "Hartree-Fock delta term"~\cite{rybak_jez_szal:1991}, $\delta_\mathrm{HF}$:
	\beq
		E_\mathrm{int}^{\mathrm{SAPT}\delta} = E^{(1)} + E^{(2)} + \delta_\mathrm{HF}.
		\label{eq:SAPTd_int}
	\eeq
It includes all leading electrostatic, exchange, induction, and dispersion contributions, all evaluated at the DFT level of theory. It should be stressed, however, that the $\delta_\mathrm{HF}$ correction comes from HF rather than DFT calculations. It provides a rough approximation of higher than second-order induction-with-exchange terms which are necessary to correct otherwise divergent perturbation expansion. More explicitly, the HF supermolecular interaction energy and~$\delta_\mathrm{HF}$ satisfy
	\beq
		\begin{split}
			E_\mathrm{int}^\mathrm{HF} & = E_\mathrm{elst}^\mathrm{(10)} + E_\mathrm{exch}^\mathrm{(10)} + \\
			 & + E_\mathrm{ind,resp}^\mathrm{(20)} + E_\mathrm{exch\text{-}ind,resp}^\mathrm{(20)} + \delta_\mathrm{HF},
		\end{split}
	\eeq
where~$E_\mathrm{elst}^\mathrm{(10)}$ and~$E_\mathrm{exch\text{-}ind}^\mathrm{(10)}$ are first-order electrostatic and exchange SAPT-HF energies, and $E_\mathrm{ind,resp}^\mathrm{(20)}$ and $E_\mathrm{exch\text{-}ind,resp}^\mathrm{(20)}$ denote second-order induction and exchange-indution SAPT-HF contributions calculated within the coupled-HF formalism.~\cite{sadlej_hhf:1980, jez_mosz_szal:1994}
In this approach, all the third-order dispersion-exchange-induction effects and dispersion-induction coupling are neglected.

\section{Results and discussion}
\label{sec:res_disc}

To test the efficiency of the PBD approach we performed calculations for two general classes of non-covalent interactions: van der Waals complexes of closed-shell atoms and ions and H-bonded systems. We applied four DFT functionals of the meta-GGA PBE hierarchy: PBREV~\cite{zhang_yang:1998} and PW91~\cite{perdew_chevary_vosko_jackson_pederson_singh_fiolhais:1992} (local exchange plus correlation), and PBE0~\cite{adamo_barone:1999} and B3LYP~\cite{becke:1993, *steph_davlin_chab_frisch:1994} (a hybrid of local and exact exchange plus correlation). All calculations have been carried out with the aug-cc-pVTZ basis sets.
 
\subsection{Closed-shell atoms and ions}

We performed calculations for several diatomic systems composed of closed-shell atoms and ions. The Ar$_2$, ArNa$^+$ and ArCl$^-$ have already been shown in Ref.~\onlinecite{rajchel_zuch_szczesniak_chal:2010_prl}, here we provide additional examples of HeLi$^+$, He$_2$, and Ne$_2$ which are typical closed shells and which provide stringent test as their intramonomer dynamic correlation is demanding for electronic structure methods. These results are compared with SAPT$\delta$ and benchmark values which typically originate from high-level supermolecular calculations.

\begin{figure}[htbp]
	\bec
	\includegraphics[width = 8.6cm]{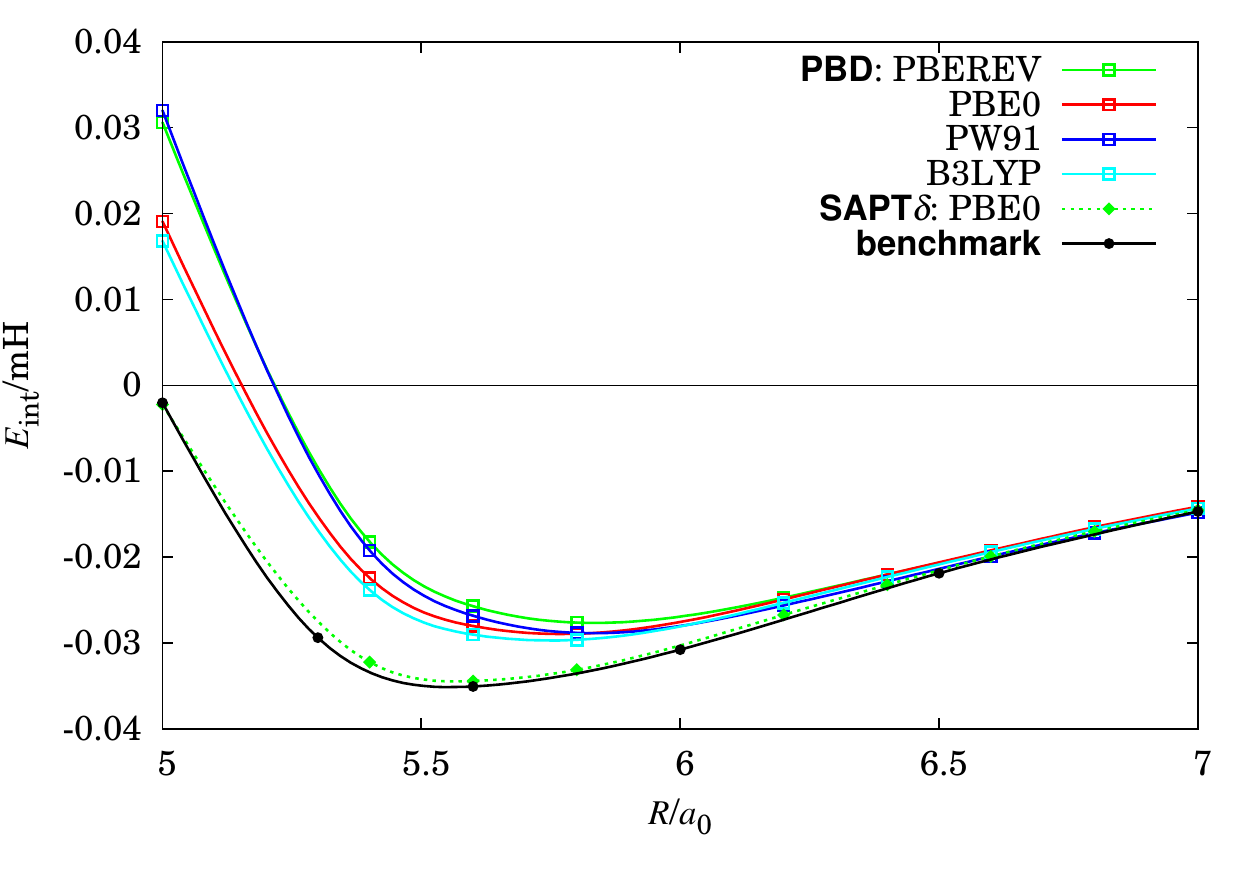}
	\caption{Comparison of PBD and SAPT$\delta$ interaction energies for~He$_2$. Benchmark results are taken from~Ref.~\onlinecite{korona_williams_bukowski_jeziorski_szalewicz:1997}.}
	\label{fig:He2_PBDSAPTd}
	\eec
	\end{figure}

\begin{figure}[htbp]
	\bec
	\includegraphics[width = 8.6cm]{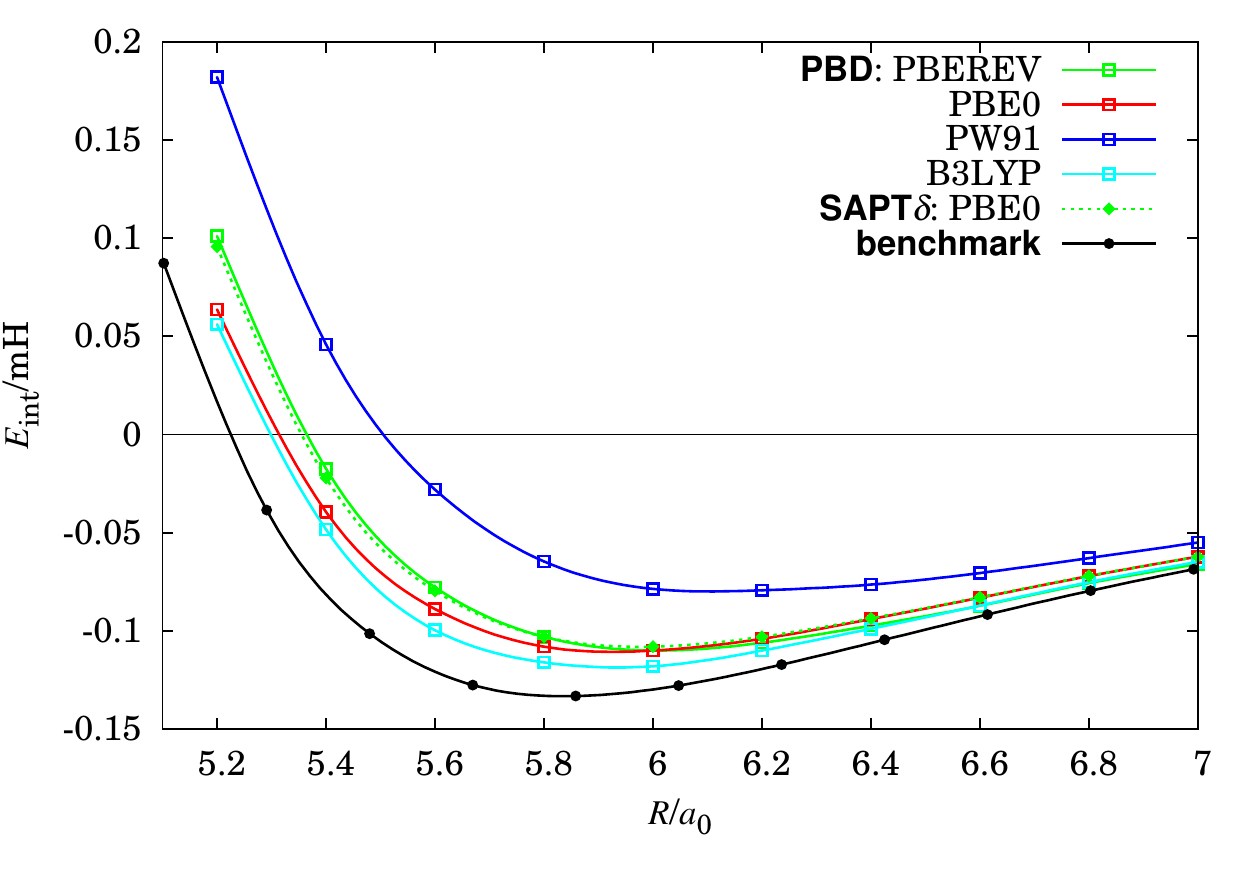}
	\caption{Comparison of PBD and SAPT$\delta$ interaction energies for~Ne$_2$. Benchmark results are taken from~Ref.~\onlinecite{hell_bich_vogel:2008}.}
	\label{fig:Ne2_PBDSAPTd}
	\eec
	\end{figure}

\begin{figure}[htbp]
	\bec
	\includegraphics[width = 8.6cm]{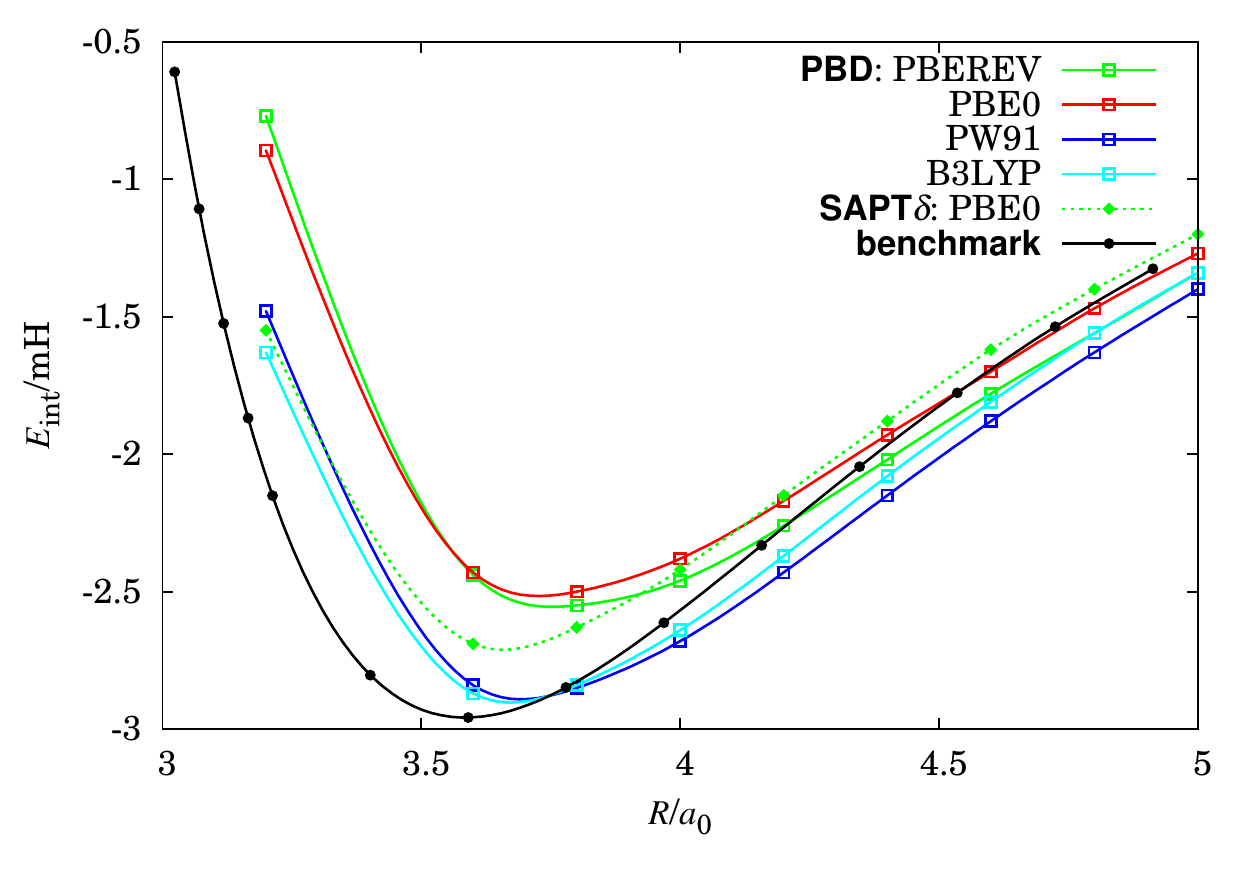}
	\caption{Comparison of PBD and SAPT$\delta$ interaction energies for~HeLi$^+$. Benchmark results are taken from~Ref.~\onlinecite{soldan_lee_lozeille_murrell_wright:2001}.}
	\label{fig:HeLi+_PBDSAPTd}
	\eec
	\end{figure}

Overall, our potentials are in good agreement with the benchmark curves for all systems under consideration, cf. Figs.~\ref{fig:He2_PBDSAPTd}--\ref{fig:HeLi+_PBDSAPTd}. As anticipated, the best performance is for the PBE0 potential. 

\begin{figure}[htbp]
	\bec
	\includegraphics[width = 8.6cm]{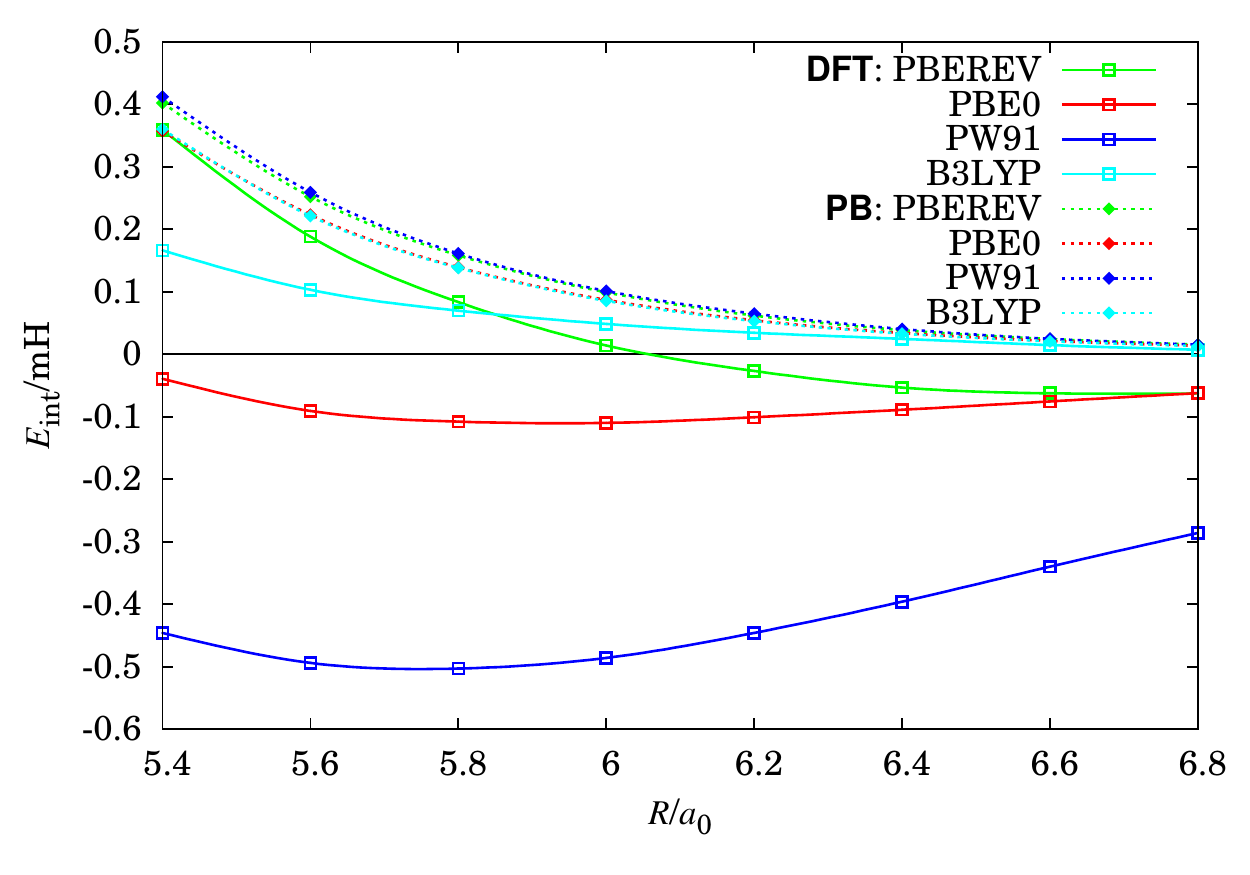}
	\caption{Comparison of PB and supermolecular DFT interaction energies for~Ne$_2$.}
	\label{fig:Ne2_PBDFT}
	\eec
	\end{figure}

One can compare the PBD interaction potentials with the potentials obtained from straightforward supermolecular calculations using the same functionals corrected for basis-set supersposition error~(BSSE). For the sake of brevity we illustrate this comparison only for Ne$_2$ (Fig.~\ref{fig:Ne2_PBDFT}), but qualitatively the results for the other two complexes (HeLi$^+$, He$_2$), as well as for the complexes in Ref.~\onlinecite{rajchel_zuch_szczesniak_chal:2010_prl} (Ar$_2$, ArNa$^+$ and ArCl$^-$) are similar. One can see in Fig.~\ref{fig:Ne2_PBDFT} that the supermolecular interaction energies reproduce neither the accurate benchmarks (they are far too shallow) nor the "dispersion-free" part as they feature a small attraction and shallow unphysical minima in the long range. Therefore, they are not appropriate for combining with with the pure dispersion term as is done in many "DFT+dispersion" approaches~\cite{elstner_hobza_frauenheim_suhai_kaxiras:2001, wu_yang:2002, zimmerli_parrinello_koum:2004, grimme:2004, chai_head-gordon:2008}. By way of contrast, the "dispersion-free" PB potentials based on the same DFT schemes (cf. Fig.~\ref{fig:Ne2_PBDFT}) are purely repulsive, revealing proper asymptotic exponential behavior, and thus can be adequately corrected by adding the dispersion contribution.

\subsection{Hydrogen-bonded and other molecular complexes}

A set of hydrogen bonded systems from~\citet{boese_martin_klopper:2007}, \citet{jur_cerny_hobza:2006} and \citet{halk_klop_helg_jorg_taylor:1999} supplemented with two typical van der Waals molecular complexes, the methane and ethylene dimers, have been used as a testing set. The selection covers a wide range of qualitatively different interactions: predominantly $\pi$-$\pi$, H-bonds with a large dispersion component (ammonia and HCl dimers), and finally strong H-bonds involving shared proton characterized by a relatively large induction contribution. The PBE0 functional and aug-cc-pVTZ basis set have been used.

A comparison between perturbation and supermolecular results requires a consistent treatment of monomer geometry in both approaches. In our comparison we performed PBD and SAPT$\delta$ calculations between monomers in their optimal dimer geometries. The databases of supermolecular interaction potentials~\cite{boese_martin_klopper:2007, jur_cerny_hobza:2006} include the effects of monomer deformations within interaction energies. For a suitable comparison with the PBD and SAPT$\delta$ data we recomputed the monomer deformation effects at the CCSD(T) level from monomer geometries of Ref.~\onlinecite{boese_martin_klopper:2007} and subtracted them from the supermolecular interaction energies. These energies are reported as benchmark in Table~\ref{tab:en_int_Hb}.

The results are shown in Table~\ref{tab:en_int_Hb}. Both the total interaction energies as well as the components are displayed, and compared with calculations by the SAPT method, and the benchmark values from appropriate references (denoted in the rightmost column of Table~\ref{tab:en_int_Hb}). The total induction energy is a sum of induction and exchange-induction contributions,
	\beq
		E_\mathrm{ind\text{-}tot}^{(2)} = E_\mathrm{ind}^{(2)} + E_\mathrm{exch\text{-}ind}^{(2)}.
	\eeq
PBD agrees favourably with benchmark results and with SAPT$\delta$. In certain classes of interactions (see discussion below) PBD yields more attractive interaction energies than both benchmark and SAPT$\delta$.

The results in Table~\ref{tab:en_int_Hb} allow one to compare the total induction effect in the PB method with that in SAPT$\delta$. To do this, $\mathscr{E}_\mathrm{def}^\mathrm{PB}$ can be compared with the sum:~$E_\mathrm{ind\text{-}tot}^{(2)} + \delta_\mathrm{HF}$. The PB induction is consistently more attractive than that from the SAPT$\delta$. The discrepancies appear to be the largest for the strong H-bonds with a shared proton, such as OH$^-$--H$_2$O, as well as ammonia-hydrogen halides where the induction effect leads to a significant stretching of the proton donor.

Table~\ref{tab:en_AC_Hb} allows us to investigate the need for the asymptotic correction~(AC). A contribution of AC to PB and SAPT energy components,
	\beq
		\delta E = \frac{E_\mathrm{AC} - E}{E_\mathrm{AC}} \cdot 100\%,
	\eeq
where $E_\mathrm{AC}$ and~$E$ denote energies calculated with and without AC, is shown in Table~\ref{tab:en_int_Hb}. For all calculations throughout the Paper the correction scheme of~\citet{grun_grits_gisb_baer:2001} have been used. It is interesting that~$\mathscr{E}_\mathrm{int}^\mathrm{HL}$ is much less sensitive than~$E^{(1)}$ to the presence of AC. This can be explained in view of the fact that both~$\mathscr{E}_\mathrm{int}^\mathrm{HL}$ and~$E^{(1)}$ involve the mutual cancellation of the electrostatic and exchange effects of which only the exchange may have a wrong asymptotic behaviour.  This attests a robust character of~$\mathscr{E}_\mathrm{int}^\mathrm{HL}$.
\begin{table*}
\caption{Comparison of interaction energies and their components from PBD calculations with SAPT(DFT) and benchmark values~(in~mH). Equilibrium geometries are from indicated references. PBE0 potentials with aug-cc-pVTZ basis set are used for monomers.}
\label{tab:en_int_Hb}
\bec
\begin{tabular}{c*{7}{D{.}{.}{2}}>{$}r<{$}}
\hline \hline
System & \mathscr{E}_\mathrm{int}^\mathrm{HL} & E^{(1)} & \mathscr{E}_\mathrm{def}^\mathrm{PB} & E_\mathrm{ind\text{-}tot}^{(2)} & \delta_\mathrm{HF} & E_\mathrm{int}^\mathrm{PBD} & E_\mathrm{int}^{\mathrm{SAPT}\delta} & $Benchmark$ \\
\hline
NH$_4^+$--H$_2$O & 
	-9.68 & -9.11 & -19.36 & -12.19 & -6.22 & -35.02 & -33.49 
 & -33.3$~\cite{boese_martin_klopper:2007}$ \\
H$_3$O$^+$--H$_2$O & 
	23.60 & 25.19 & -103.72 & -56.14 & -38.37 & -93.26 & -82.46 
 & -83.0$~\cite{boese_martin_klopper:2007}$ \\
OH$^-$--H$_2$O & 
	11.70 & 13.02 & -61.19 & -29.65 & -21.53 & -63.77 & -52.43 
 & -52.4$~\cite{boese_martin_klopper:2007}$ \\
(H$_2$O)$_2$ & 
	-1.02 & -0.89 & -3.92 & -2.16 & -1.45 & -8.56 & -8.12 
 & -9.00$~\cite{jur_cerny_hobza:2006}$ \\
(HF)$_2$ & 
	-0.69 & -0.05 & -4.08 & -2.43 & -1.19 & -7.59 & -6.48 
 & -7.22$~\cite{halk_klop_helg_jorg_taylor:1999}$ \\
(HCl)$_2$ & 
	 1.65 &  1.77 & -2.20 & -1.02 & -0.98 & -3.48 & -3.16 
 & -3.10$~\cite{halk_klop_helg_jorg_taylor:1999}$ \\
(NH$_3$)$_2$ & 
	-0.48 & -0.11 & -1.55 & -0.84 & -0.52 & -5.34 & -4.78 
 & -5.05$~\cite{jur_cerny_hobza:2006}$ \\
NH$_3$--H$_2$O & 
	-0.15 &  0.23 & -6.61 & -3.26 & -2.44 & -11.50 & -10.21 
 & -10.3$~\cite{boese_martin_klopper:2007}$ \\
H$_3$N--HF & 
	 1.08 &  2.47 & -18.17 & -9.30 & -6.41 & -23.94 & -20.09 
 & -20.8$~\cite{boese_martin_klopper:2007}$ \\
H$_3$N--HCl & 
	 8.35 &  9.05 & -18.53 & -7.32 & -8.59 & -18.28 & -14.96 
 & -14.3$~\cite{boese_martin_klopper:2007}$ \\
(CH$_4$)$_2$ & 
	 0.70 &  0.58 & -0.07 & -0.01 & -0.04 & -0.85 & -0.95 
 & -0.81$~\cite{jur_cerny_hobza:2006}$ \\
(C$_2$H$_4$)$_2$ & 
	 1.71 &  1.35 & -0.34 & -0.11 & -0.23 & -2.11 & -2.47 
 & -2.58$~\cite{jur_cerny_hobza:2006}$ \\
\hline \hline
\end{tabular}
\eec
\end{table*}

\begin{table*}
\caption{Comparison of asymptotic corrections~(in~\%) for PB and SAPT components of interaction energies.}
\label{tab:en_AC_Hb}
\bec
\begin{tabular}{c*{4}{D{.}{.}{2}}}
\hline \hline
System & \delta \mathscr{E}_\mathrm{int}^\mathrm{HL} & \delta E^{(1)} & \delta \mathscr{E}_\mathrm{def}^\mathrm{PB} & \delta E_\mathrm{ind\text{-}tot}^{(2)} \\
\hline
NH$_4^+$--H$_2$O & 
	 8.8 & 16.2 & -4.2 & -0.5 
 \\
H$_3$O$^+$--H$_2$O & 
	-7.5 & -8.8 & -1.7 & -1.0 
 \\
OH$^-$--H$_2$O & 
	 5.2 &  1.0 &  0.3 & -1.6 
 \\
(H$_2$O)$_2$ & 
	27.0 & 124.5 & -7.8 &  0.8 
 \\
(HF)$_2$ & 
	21.8 & 857.0 & -4.0 & -0.2 
 \\
(HCl)$_2$ & 
	-2.9 & -11.7 & -3.0 &  0.4 
 \\
(NH$_3$)$_2$ & 
	 3.2 & 158.8 & -1.8 &  0.1 
 \\
H$_3$N--HF & 
	-3.3 & -14.2 & -0.9 & -0.7 
 \\
(CH$_4$)$_2$ & 
	-0.9 & -41.8 &  5.9 & 24.9 
 \\
(C$_2$H$_4$)$_2$ & 
	-5.7 & -54.2 & -21.4 & -4.5 
 \\
\hline \hline
\end{tabular}
\eec
\end{table*}

\begin{figure}[htbp]
	\bec
	\includegraphics[width = 8.6cm]{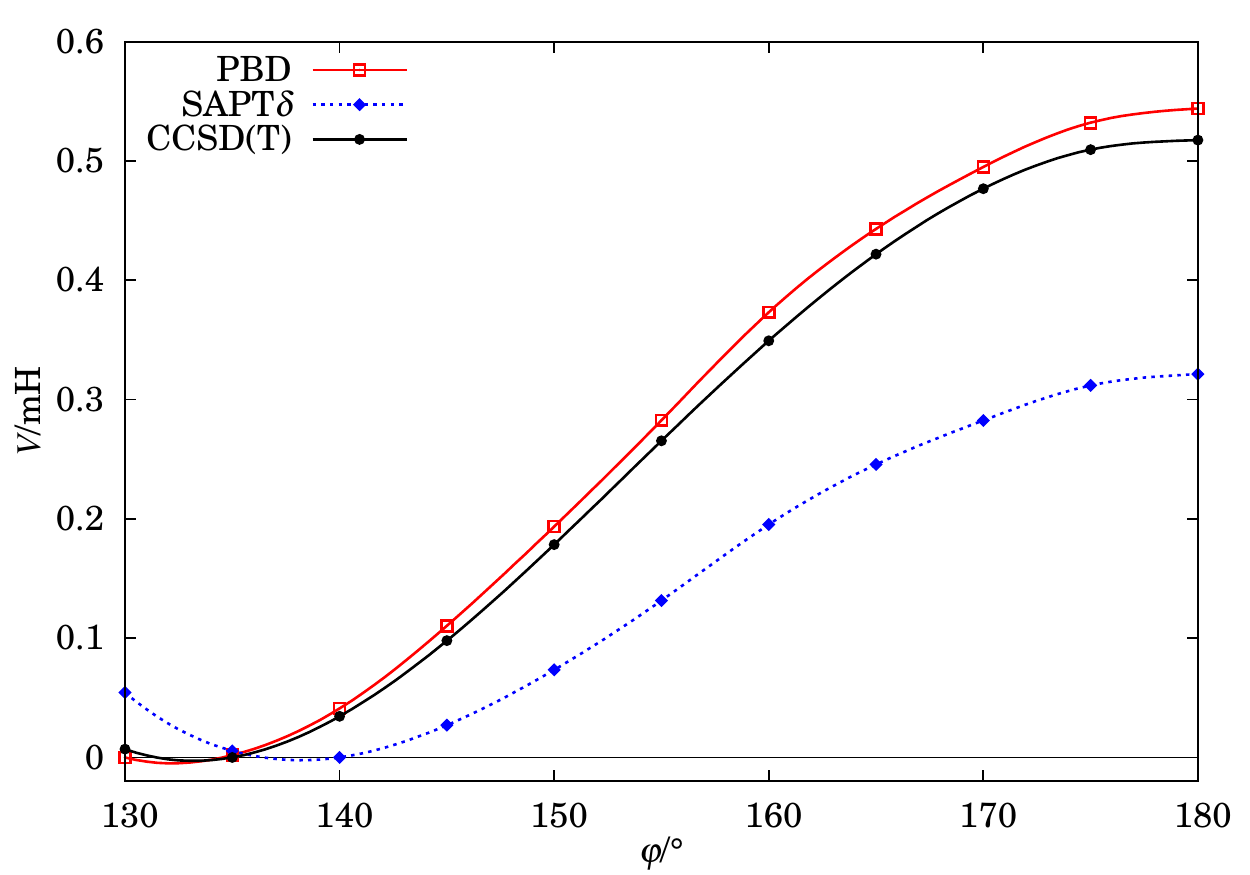}
	\caption{Comparison of interaction energies for H$_2$O--HF dimer with respect to the bending angle.}
	\label{fig:H2O-HF_bend}
	\eec
	\end{figure}

Another test has been carried out for the bending potential of the H$_2$O--HF complex. The water and HF geometries and the intersystem distance were taken from Ref.~\onlinecite{halk_klop_helg_jorg_taylor:1999}. The potential~$V(\varphi)$, where $\varphi$~is the angle between water plane and the line connecting oxygen and HF hydrogen atoms, is the interaction energy scaled so that it is zero at the minimum. As can be seen in Fig.~\ref{fig:H2O-HF_bend}, the PBD method gives a remarkably similar potential to CCSD(T), while the SAPT$\delta$ underestimates the barrier by about~40~\%.

\section{Summary and conclusions}

A new DFT approach to calculations of van der Waals complexes has been derived, and tested for a variety of systems ranging from noble gas dimers to different, weakly and strongly H-bonded dimers. 

The new formalism is based on the concept of interacting separated monomers~\cite{rajchel_zuch_szczesniak_chal:2010_cpl}. The monomers are described within the DFT formalism. Their densities interact under the constraint of the antisymmetry principle and under the exact exchange intermolecular potential until self-consistency is reached. The resulting interaction energy represents the dispersion-free part of the total interaction energy. The new formalism provides a consistent definition of the "non-dispersion" part of the interaction energy at the DFT level of theory. The total interaction energy is obtained by \textit{a posteriori} adding the DFT dispersion contribution from SAPT or other formalisms. It should be stressed that the PB treatment \emph{does not} require any extra empirical and adjustable parameters (besides those that are already used by the DFT description of monomers). 

It has been demonstrated that for rare gas dimers, hydrocarbon dimers, and both weak and strong H-bonded dimers including ionic interaction, the PB combined with monomer description of PBE0 provides results that agree well with the benchmark values.

The PB technique may be used for clusters of atoms and/or molecules and is not restricted to closed-shell systems. It may be combined with different functionals to describe the monomers. Similarly, it may be combined with a variety of formalisms to calculate the dispersion part. 

\section{Acknowledgments}

Financial support from NSF (Grant~No.~CHE-0719260) is gratefully acknowledged. We acknowledge computational resources purchased through NSF~MRI program (Grant~No.~CHE-0722689).


\bibliography{Bibliography}

\end{document}